\documentclass{ws-procs975x65}

\begin{document}

\title{EVOLUTION OF A VOID AND AN ADJACENT GALAXY SUPERCLUSTER IN 
THE QUASISPHERICAL SZEKERES MODEL}

\author{Krzysztof Bolejko}

\address{Nicolaus Copernicus Astronomical Center, Polish Academy of Sciences,\\
ul. Bartycka 18, 00-716 Warsaw, Poland\\
\email{bolejko@camk.edu.pl}}

\begin{abstract}
This paper investigates the evolution 
of a void and an adjacent galaxy supercluster.
For this purpose the quasispherical Szekeres model is employed. The Szekeres model
is an exact solution of the Einstein field equations.
In this way investigations of the evolution of the cosmic structures presented
here can be freed from such assumptions
as small amplitude of the density contrast.
Studying the evolution of a void and an adjacent supercluster  
results not only in better understanding the evolution
of cosmic structures but also in acquainting us with the Szekeres model.
 The main results include the conclusion that 
 small voids surrounded by large overdensities evolve slower than large, 
isolated voids do. On the other hand, large voids enhance the evolution 
of adjacent superclusters which evolve much faster than isolated galaxy superclusters.
\end{abstract}


\bodymatter

\section{Introduction}
The structures which can be observed in the local Universe include small voids among compact clusters, superclusters and large voids surrounded by large walls or long filaments.
 The present day density contrast of 
overdense regions is larger than 1\cite{BE}
and inside voids it descends to -1\cite{Hoy}. 
To describe these structures, an exact solution of the Einstein field equations must be
employed. However, among the known exact solutions none is flexible enough
to describe such complicated structure as our Universe.
Therefore, the analysis of this paper will focus on smaller scales. The structures on small scales, up to Mpc
can be described by the Szekeres model which is an exact solution
of the Einstein field equations.

The structure of this paper is as follows: Sec.
\ref{szekmdl} presents the Szekeres model;
Sec. \ref{doubstr} presents the evolution of 
pairs void--supercluster in the quasispherical Szekeres model;
Sec. \ref{expan} presents the role of expansion in
the process of structure formation.

\section{The Szekeres model}\label{szekmdl}

For our purpose it is convenient to use a coordinate system different from
that in which Szekeres\cite{Sz1} originally found his solution. The metric is of
the following form\cite{HK}:

\begin{equation}
ds^2 =  c^2 dt^2 - \frac{(\Phi' - \Phi \frac{\textstyle E'}{\textstyle E})^2}
{(\varepsilon - k)} dr^2 - \Phi^2 \frac{(dp^2 + dq^2)}{E^2}, \label{ds2}
 \end{equation}
where ${}' \equiv \partial/\partial r$, $\varepsilon = \pm1,0$ and $k = k(r)
\leq \varepsilon$ is an arbitrary function of $r$.

The function $E$ is given by: 
 \begin{equation}
E(r,p,q) = \frac{1}{2S}(p^2 + q^2) - \frac{P}{S} p - \frac{Q}{S} q + C ,
 \end{equation}
where the functions $S = S(r)$, $P = P(r)$, $Q = Q(r)$, and $C = C(r)$ satisfy
the relation:
 \begin{equation}
C = \frac{P^2}{2S} + \frac{Q^2}{2S} + \frac{S}{2} \varepsilon,~~~~~~~~~
\varepsilon = 0, \pm 1,
 \end{equation}
but are otherwise arbitrary.

The quasispherical Szekeres model is the case of $\varepsilon = 1$.
As was shown in Ref.~\refcite{Sz2}, surface of  $t$ = const and $r$ = const
has a topology of a sphere. However, as
$S, P$ and $Q$ are now functions of $r$, the spheres are not concentric. For the
spheres to be concentric, the functions $S,P$ and $Q$ must be constant.
Such conditions entail spherical symmetry, with which  the Szekeres model 
becomes the the Lema\^itre--Tolman model\cite{Lem,Tol}.

Substituting the metric (\ref{ds2}) in the Einstein equations, and assuming the
energy momentum tensor for a dust, the Einstein equations reduce
 to the following two:

\begin{equation}
\frac{1}{c^2} \dot{\Phi}^2 (t,r) = \frac{2M(r)}{\Phi(t,r)} - k(r) + \frac{1}{3} \Lambda
\Phi^2(t,r), \label{vel}
\end{equation}

\begin{equation}
4 \pi \frac{G}{c^2} \rho(t,r,p,q) = 
 \frac{M'(r) - 3 M(r) E'(r,p,q)/E(r,p,q)}{\Phi^2(t,r) [ \Phi'(t,r) - \Phi(t,r) E'(r,p,q)/E(r,p,q)]}. \label{rho}
\end{equation}

Equation (\ref{vel}) can be integrated:

\begin{equation}
\int\limits_0^{\Phi}\frac{{\rm d}
\tilde{\Phi}}{\sqrt{\frac{2M(r)}{\tilde{\Phi}} - k(r) + \frac{1}{3} \Lambda
\tilde{\Phi}^2}} = c \left[t- t_{\mathrm{B}}(r)\right]. \label{cal}
\end{equation}

As can be seen the Szekeres model is specified by 6 functions: 
$M(r), k(r), t_{\mathrm{B}}(r), S(r), Q(r), P(r)$.
However, by a
choice of the coordinates, the number of independent functions can be reduced
to 5.

The equations of motion $T^{\alpha \beta}{}_{; \beta} = 0$ are reduced to the
continuity equation:

\begin{equation}
\dot{\rho} + \rho \Theta = 0,
\label{coneq}
\end{equation}
where $\Theta$ is the scalar of expansion and is equal to:

\begin{equation}
\Theta(t,r,p,q) = 
3 \frac{\dot{\Phi}(t,r)}{\Phi(t,r)} + \frac{ \dot{\Phi}'(t,r) - 
\dot{\Phi}(t,r) \Phi'(t,r)/\Phi(t,r)}{\Phi'(t,r) - \Phi(t,r) E'(r,p,q)/E(r,p,q)}. \label{eks}
\end{equation}

In the expanding Universe $\Theta$ is positive
so the density decreases. The structures which  exist in the Universe,
emerged either due to slower expansion of the space (formation of 
overdense regions) or 
due to faster expansion (formation of underdense regions).
In the Friedmann limit $R \rightarrow r a$, where $a$ is the scale factor and $\Theta \rightarrow 3 H_0$.

The Szekeres model is known to have no symmetry\cite{BST}. It is of great flexibility and wide application in cosmology\cite{BT} and in astrophysics\cite{Sz2,HK}, and still it can be used as a model of many astronomical phenomena. 
In this paper it will be employed to study the evolution of cosmic structures in 
different environments.

\subsection{Density contrast}

To compare the evolution of different models the change in their
density contrast will be  considered.
Two different types of density contrast indicators
are taken into account. The first one is the usual density contrast,
$\delta = \rho / \rho_{\mathrm{b}} - 1$, the second one 
is the spatially invariant density contrast\cite{MT}:

\begin{equation}
S_{\mathrm{IK}} = \int_{\Sigma} \left| \frac{h^{\alpha \beta}}{\rho^I} \frac{\partial \rho}{ \partial x^{\alpha}}
\frac{\partial \rho}{ \partial x^{\beta}} \right|^K {\rm d} V,
\end{equation}
where $I \in \mathbb{R}$, and $K \in \mathbb{R} \backslash \{0\}$.
This family of the density contrast indicators can be
considered as local or global depending on the size of $\Sigma$.
Such a quantity not only describes the change of density
but also the change of gradients and the volume
of a perturbed region. So this density indicator 
describes the evolution of the whole region
in a more sophisticated way than the  $\delta$.
Here only the case $I=2, K=1/2$ will be considered.

\subsection{Model set--up}

To specify the model 5 functions
of the radial coordinate
need to be known. Let us define the radial coordinate as a
value of $\Phi$ at the initial instant $t_0 =  0.5$ My after the big bang,
i.e., $r:= \Phi(r,t_0)$.

Two of these functions will be $t_{\mathrm{B}}(r)$ and $M(r)$.
 Let us write the mass function in the following form:

\begin{equation}
M(r) = M_0(r) + \delta M(r),
\end{equation}
where $M_0$ is the mass distribution as in the homogeneous universe,
and $\delta M$ is a mass correction, which can be either positive or negative.
The $\delta M$ is defined similarly as in the spherically symmetric case:

\begin{equation}
\delta M(r) = 4 \pi \frac{G}{c^2} \int_0^r {\rm d} \tilde{r} R(\tilde{r},t_0)^2 R'(\tilde{r},t_0) 
\delta \bar{\rho}(\tilde{r}),
\end{equation}
where $\delta \bar{\rho}(r)$ is an arbitrary function chosen to specify the $\delta M$.
Although $\delta \bar{\rho}(r)$ is not the initial function of density fluctuations (since an initial density fluctuation is a function of all coordinates) it gives some estimation on the initial density fluctuation of the 
monopole density component.

The next three functions are $P(r), Q(r), S(r)$.
 All functions defining the model are presented 
in Table \ref{Tab1}.
The numerical algorithm used to solve the Szekeres model's equations
is presented in detail in Ref.~\refcite{KBysc}.

The chosen background model is the homogeneous Friedmann model with the density:
 \begin{equation}
   \rho_{\mathrm{b}} = \Omega_{\mathrm{m}} \times \rho_{\mathrm{cr}} = 0.24 \times \frac{3H_0^2}{8 \pi G}.
   \label{rbdf}
 \end{equation}
 where the Hubble constant is $H_0 =74$ km s$^{-1}$ Mpc$^{-1}$.  The cosmological constant,  $\Lambda$, corresponds to $\Omega_{\Lambda} = 0.76$, where $\Omega_{\Lambda} = (1/3)  ( c^2 \Lambda/H_0^2)$.

\section{Models of a  void and an adjacent galaxy supercluster}\label{doubstr}

In this section the evolution a void with an adjourning galaxy supercluster is investigated.
Although within the Szekeres model 
more than two structures can be described,
such investigations of less complex cases are useful
because they enable us to draw some general conclusions
without going into too much detail, which could easily
obscure the larger picture.

\subsection{Models with $P' = 0 = S',~Q' \ne 0$}\label{p's'z}

As mentioned above, if $P' = 0 = S' = Q'$ the Szekeres model becomes
the Lema\^itre--Tolman model. Hence, the class of models considered 
in this subsection is the simplest generalisation of 
the spherically symmetric models.

The double structure of a void and adjourning supercluster 
can be described in the Szekeres model in two different ways.
The first alternative is when $\delta M<0$, the second when $\delta M>0$.
Both these possibilities are examined here.

\subsubsection{Model specification}

The exact form of the functions used to define models 1 and 2 is presented in Table \ref{Tab1}. 
The density distributions of models
1 and 2 are presented in Fig. \ref{fig1}. As can be seen
the model with $\delta M <0$ has the void in the center, and the adjacent 
supercluster has an elongated shape. 
It is the opposite in model 2. The overdense region at the origin 
is more compact than in model 1, and the adjacent element is the void.

\begin{table}
\tbl{The exact form of the functions used to define models 1--4.}
{\begin{tabular}{@{}cccccc@{}}
\toprule
Model & $t_{\mathrm{B}}$ & $\delta \bar{\rho}$ & $S$ & $P$& $Q$ \\
\colrule
1 & 0 &$-5 A \times \exp[-(\ell/8)^2]$ & 1 & 0 & $b \ln (1+\ell) \times \exp (-3 A \ell)$ \\
2 & 0 & $1.14 A \times \exp[-(\ell/9)^2]$ & 1 & 0 & $c \ln (1+ 0.2 {\ell})   \times \exp (-3 A \ell)$ \\
3 &  0 & $-5 A \times \exp[-(\ell/8)^2]$ & $-\ell^{0.4}$ & $0.55 \ell^{0.4}$ & $0.33 \ell^{0.4}$\\
4 &  0 & $1.14 A \times \exp[-(\ell/9)^2]$ & $-\ell^{0.9}$ & $0.55 \ell^{0.8}$ & $0.33 \ell^{0.8} $\\
\botrule
\end{tabular}
}
\begin{tabnote}
$\ell = r/{\rm Kpc}$, $A = 10^{-3}$, $b=-0.6$, $c = 1.45$,.\\
\end{tabnote}
\label{Tab1}
\end{table}

\def\figsubcap#1{\par\noindent\centering\footnotesize(#1)}

\begin{figure}%
\begin{center}
  \parbox{2.1in}{\epsfig{figure=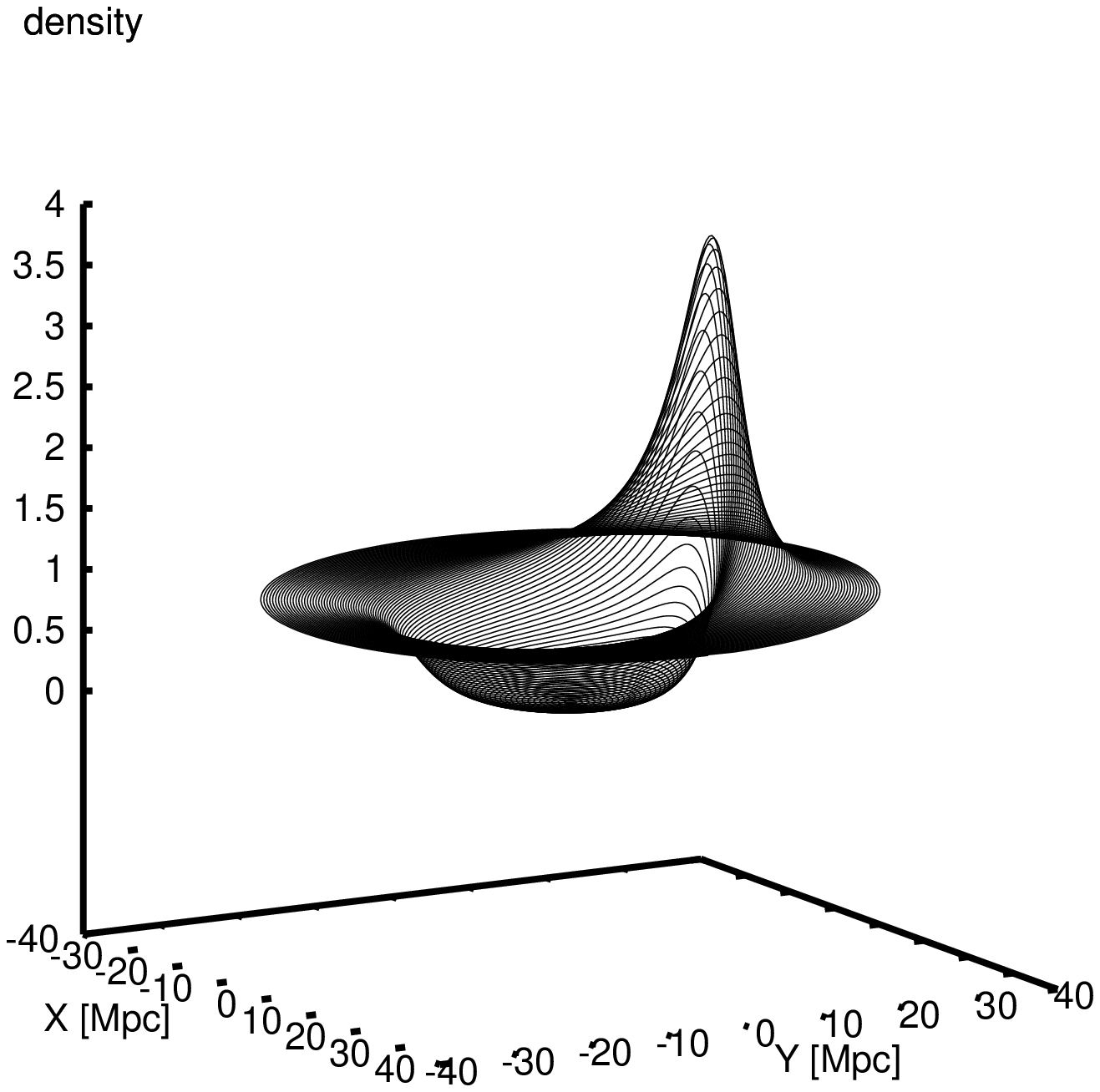,scale=0.38}\figsubcap{a}}
  \hspace*{4pt}
  \parbox{2.1in}{\epsfig{figure=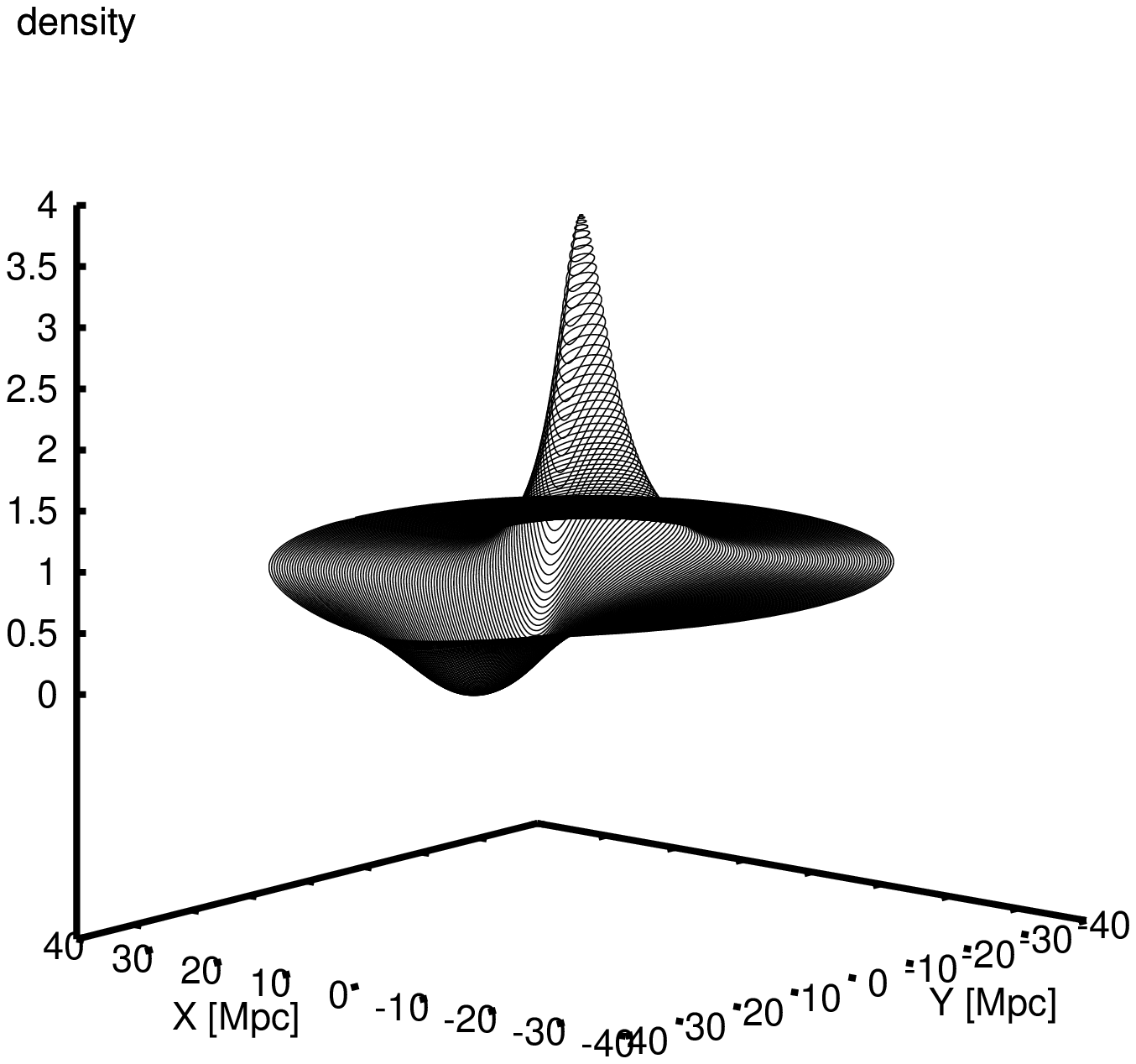,scale=0.38}\figsubcap{b}}
  \caption{The present--day density distribution, $\rho/\rho_{\mathrm{b}}$.
Fig. 1(a) presents model 1 ($\delta M<0$). Fig. 1(b) presents model 2 ($\delta M>0$).}%
  \label{fig1}
\end{center}
\end{figure}

\subsubsection{Evolution}\label{evoqnez}

In this section we compare the evolution of the density contrast, $\delta(t,r,\theta, \phi)$, and the 
$S_{2,1/2}(t, \Sigma)$ density indicator   for models 1,  2, with the corresponding models of a single void and the models of a single supercluster
obtained within the Lema\^itre--Tolman model.
The Lema\^itre--Tolman model is considered because within this model
one can describe a single  spherically symmetric structure.
Such a comparison can demonstrate
how the evolution of a structure
changes if there is another structure in its close neighbourhood.

\begin{figure}%
\begin{center}
  \parbox{2.1in}{\psfig{file=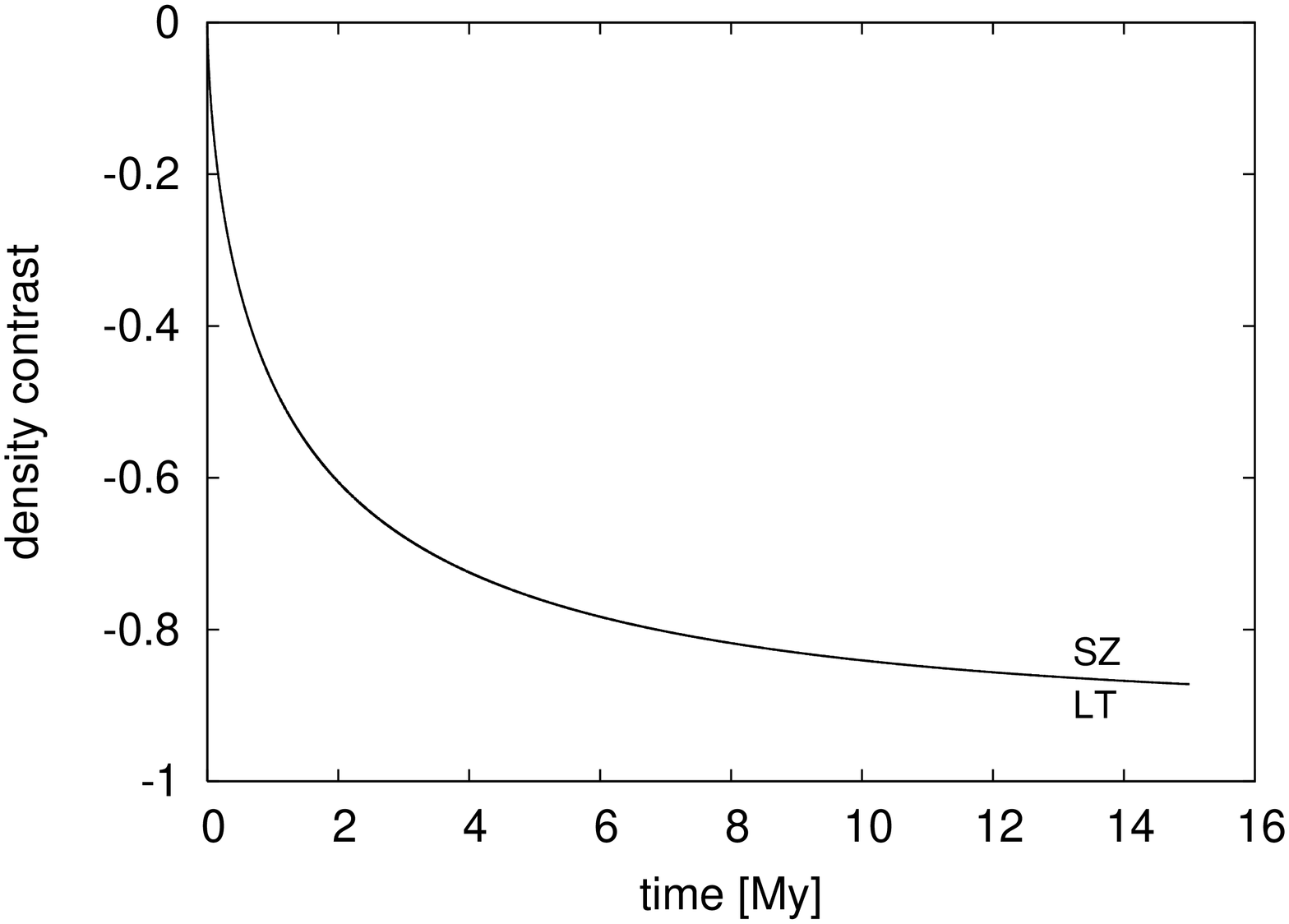, scale=0.23, angle=270}\figsubcap{a}}
  \hspace*{4pt}
  \parbox{2.1in}{\psfig{file=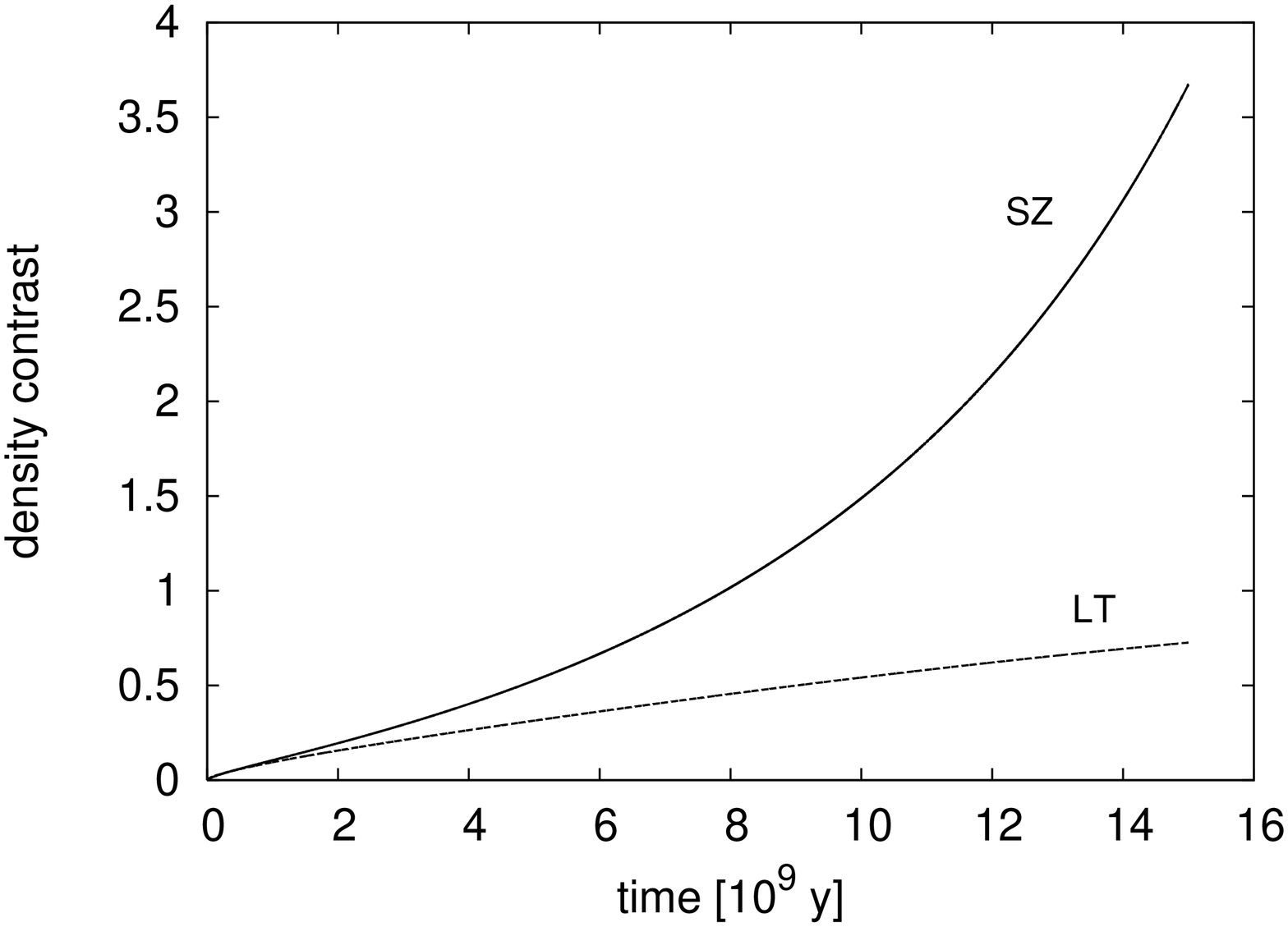, scale=0.23, angle=270}\figsubcap{b}}
  \caption{The evolution of the density contrast inside the void (a), and inside the supercluster (b) for model 1 ($\delta M<0$). The curve SZ presents the evolution 
within the Szekeres model; curve  LT presents the 
evolution within the Lema\^itre--Tolman model.}%
  \label{fig2}
\end{center}
\end{figure}

Fig. \ref{fig2} presents the evolution of the density contrast of model 1 in comparison with the corresponding 
models obtained within the Lema\^itre--Tolman model.
The Lema\^itre--Tolman model was specified by assuming the same condition 
as the ones in the Szekeres model at the initial instant.
The local density contrast, $\delta$, is compared at the point of the 
maximal and minimal density value.
Fig. \ref{fig2}(a) presents the evolution of the density contrast inside the void.
 As can be seen
the bahaviour of the density contrast in both models is similar.
This due to the conditions of regularity at the origin (for a detailed description of the regularity conditions at the origin see Ref.~\refcite{HK}).
These conditions imply that the origin  in the quasispherical 
Szekeres and Lema\^itre-Tolman model  behaves like a Friedmann models
Fig. \ref{fig2}(b) presents the evolution of the density contrast 
at the very center of the overdense region
of the model 1 and the corresponding Lema\^itre--Tolman model. 
The growth of density contrast in the Szekeres model is much faster than in the corresponding 
Lemaitre--Tolman model.
The results of this comparison indicate that within the perturbed region of mass below the background 
mass ($\delta M <0$) the evolution of underdensities does not change but the evolution of the overdense regions situated at the edge of the  underdense regions is much faster than
the similar evolution of isolated structures.

\begin{figure}%
\begin{center}
  \parbox{2.1in}{\psfig{file=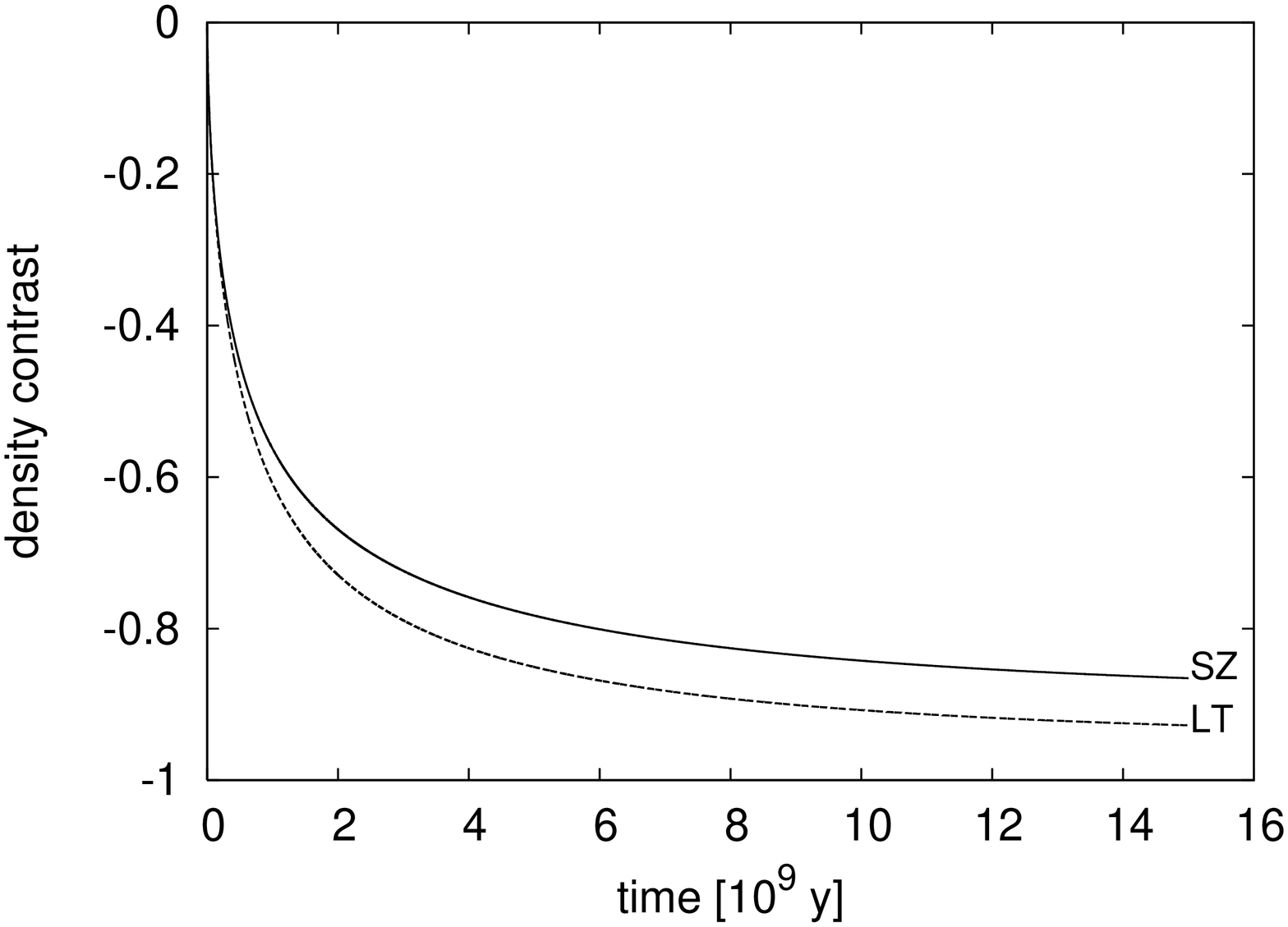, scale=0.23, angle=270}\figsubcap{a}}
  \hspace*{4pt}
  \parbox{2.1in}{\psfig{file=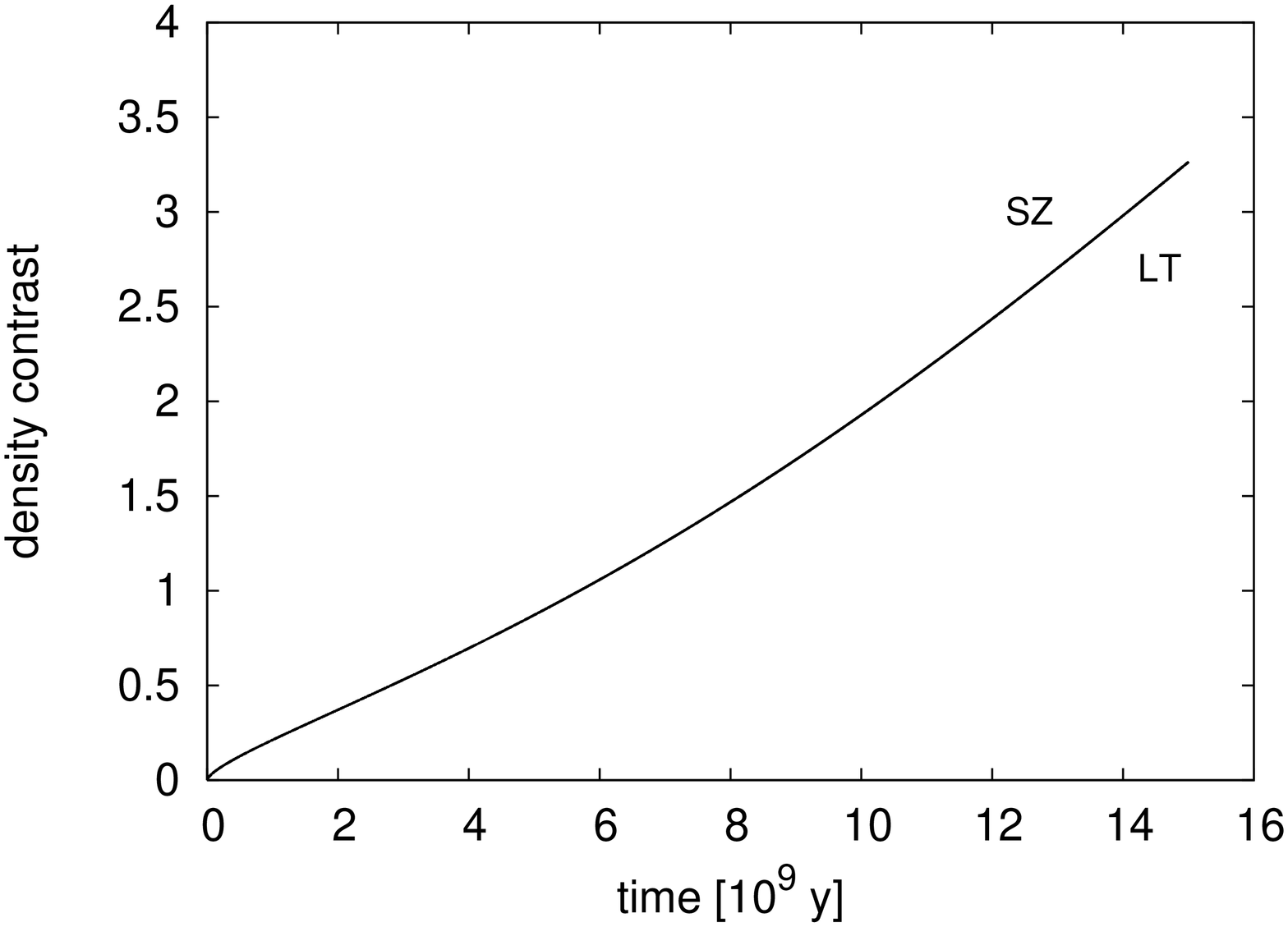, scale=0.23, angle=270}\figsubcap{b}}
  \caption{The evolution of the density contrast inside the void (a), and inside the supercluster (b) for model 2 ($\delta M>0$). The curve SZ presents the evolution 
within the Szekeres model; curve LT presents the 
evolution within the Lema\^itre--Tolman model.}%
  \label{fig3}
\end{center}
\end{figure}

The evolution of the density contrast of model 2 ($\delta M >0$) is presented 
in Fig. \ref{fig3}, the evolution of the density contrast at the point of minimal density is depicted in Fig. \ref{fig3}(a), and the evolution at the origin is depicted in Fig. \ref{fig3}(b).
Similarly as in model 1, the evolution at the origin in the Szekeres model and in the Lema\^itre--Tolman model are very much alike.
The evolution of the void, however, is slower within the Szekeres model than it is in the Lema\^itre--Tolman model.
This implies that single, isolated voids evolve much faster than the ones which are in the neighborhood
of large overdensities where the mass of the perturbed region is above the background mass ($\delta M>0$).

\subsection{Models with $P' \ne 0 \ne S',~Q' \ne 0$}\label{sqpnz}

In this section models of non--constant $P, Q$ and $S$  
are investigated.
The evolution of these models is compared with the evolution 
of models which were considered in Sec. \ref{p's'z}

\subsubsection{Models specification}

\begin{figure}%
\begin{center}
  \parbox{2.1in}{\epsfig{figure=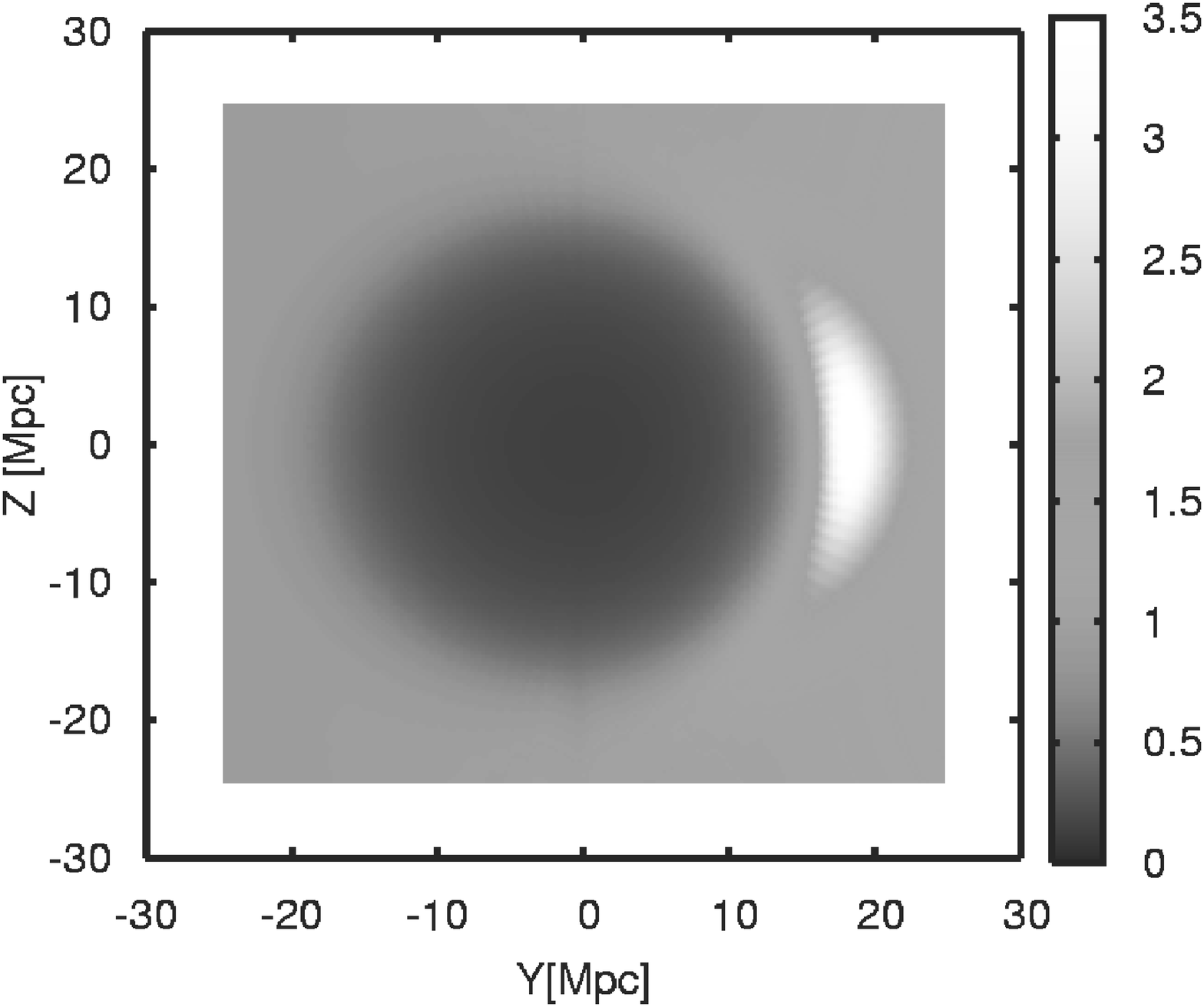, scale=0.17}\figsubcap{a}}
  \hspace*{4pt}
  \parbox{2.1in}{\epsfig{figure=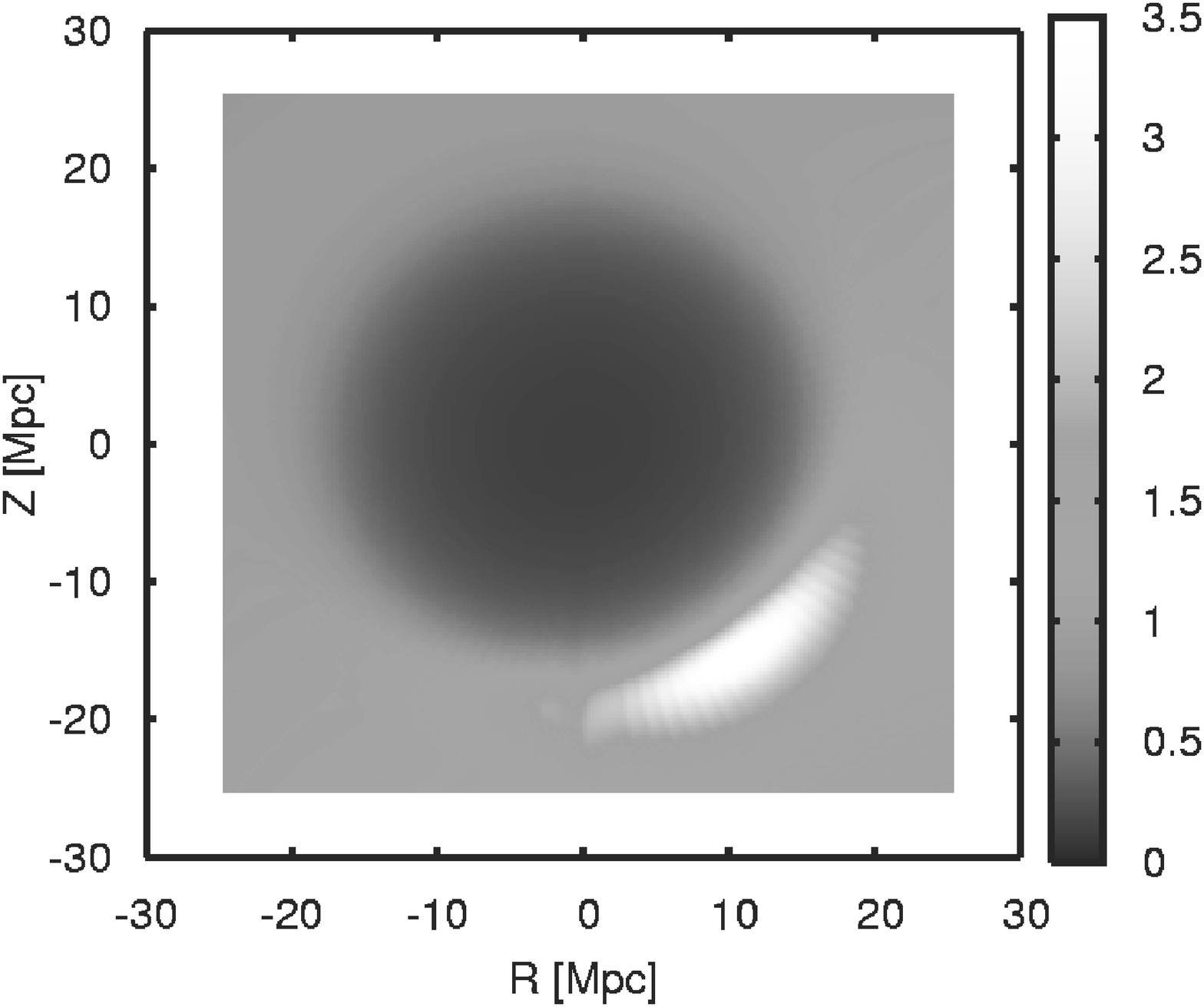, scale=0.17}\figsubcap{b}}
  \caption{The present--day colour coded density distribution, $\rho/\rho_{\mathrm{b}}$, of models with $\delta M <0$. (a)  presents the model with $P' = S' = 0$. (b) presents the model with $P' \ne 0 \ne S'$.}%
  \label{fig5}
\end{center}
\end{figure}

\begin{figure}%
\begin{center}
  \parbox{2.1in}{\epsfig{figure=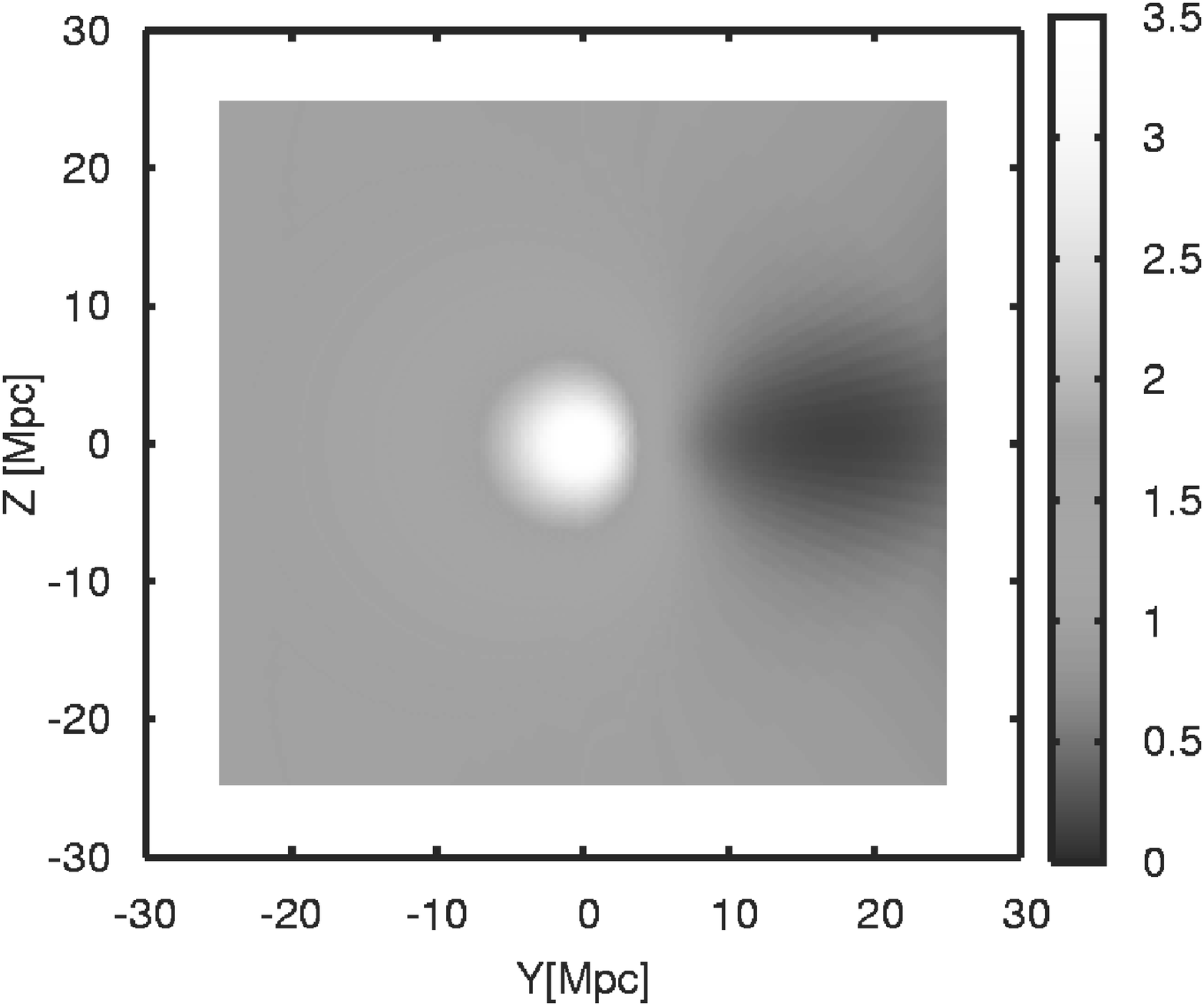, scale=0.17}\figsubcap{a}}
  \hspace*{4pt}
  \parbox{2.1in}{\epsfig{figure=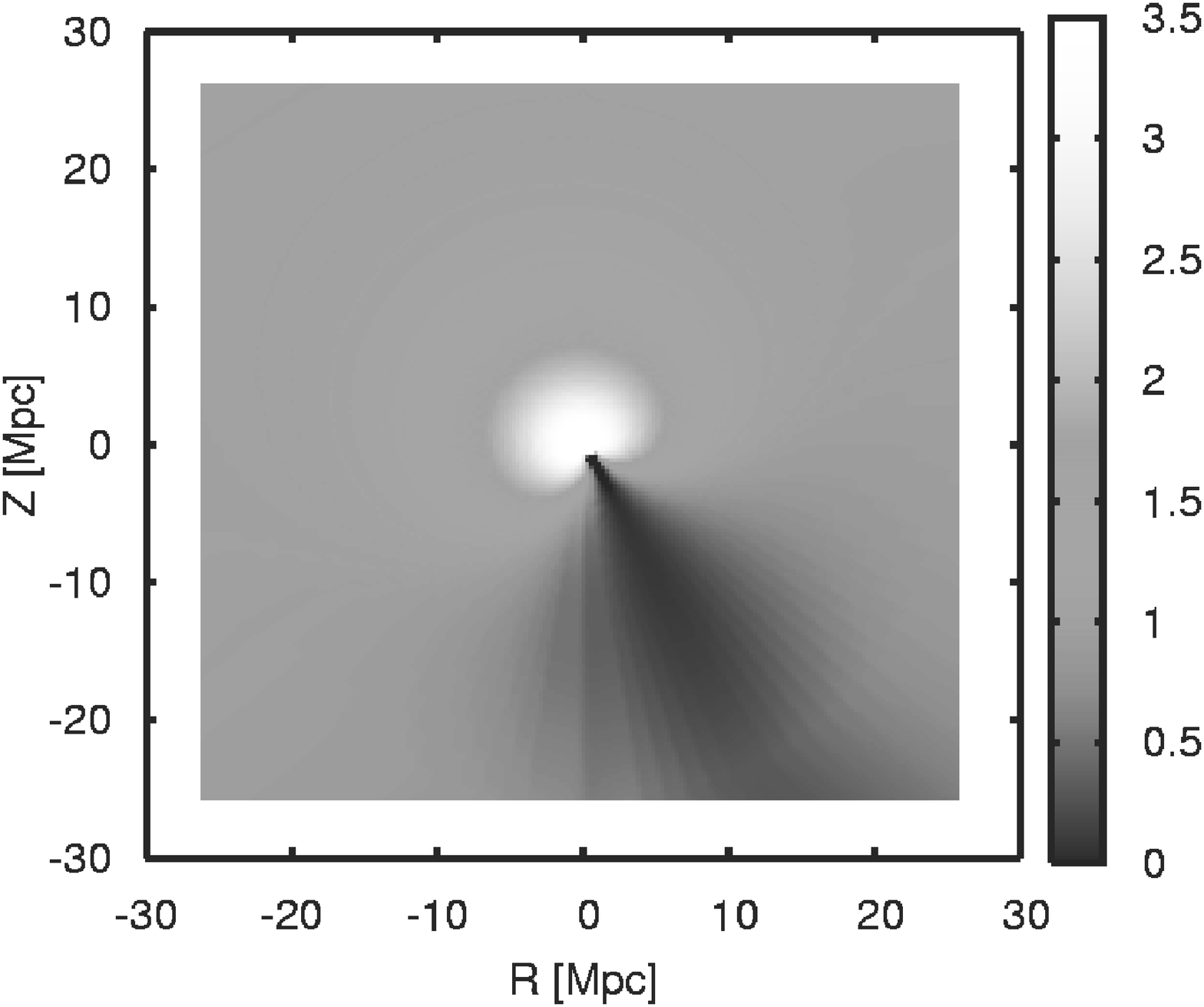, scale=0.17}\figsubcap{b}}
  \caption{The present--day colour coded density distribution, $\rho/\rho_{\mathrm{b}}$, of models with $\delta M > 0$. (a)  model with $P' = S' = 0$. (b) model with $P' \ne 0 \ne S'$.}%
  \label{fig6}
\end{center}
\end{figure}

The exact form of the functions used to define models 3 and 4 is presented in Table \ref{Tab1}. Fig. \ref{fig5} presents the comparison of the present day density distribution 
in models 1 and 3 in colour coded diagrams. It presents the vertical cross--sections of 
the considered
structures. Fig. \ref{fig5}(a) presents the vertical cross--section
through the surface of $\phi = \pi/2$ and Fig. \ref{fig5}(b)  presents
the cross section through the surface of $\phi \approx \pi/6$.
The comprehensive study of the 
 vertical and horizontal cross--sections
of similar models was presented in Ref.~\refcite{KBprd}.
Fig. \ref{fig6} also presents the vertical cross--sections 
of models 2 and 4.
As can be seen, both structures appear to be similar but,
in comparison with model 1, in model 3 the additional component is moved down and right. 
Model 4 on the other hand presents the structure moved down and right in comparison with model 2.

\subsubsection{Evolution}

The evolutions of the density contrast inside the voids and superclusters of models 3 and 1 are very similar which needn't be suprising as 
 model 3 has the same $\bar{\delta}(r)$ as model 1.
Also, the evolutions of the corresponding density contrasts of model 4 and 2
are similar. 
The functions $S,P,Q$ were chosen so they reproduce the same shape of current
structures and the same density contrast inside them --- that
is why  the evolution of a local density contrast is comparable for models 1 and 3, and for models 2 and 4.
However, it is not clear whether or not the evolution of
$S_{2,1/2}$ is comparable too.
 When the functions $S, P, Q$ are not constant,
the axis of a density dipole changes. Also, the volume of the perturbed region 
as well as the density gradients can be different. So it may be interesting to compare the evolution of the whole perturbed underdense and overdense regions of models 1, 2, 3, and 4.

\begin{figure}%
\begin{center}
  \parbox{2.1in}{\epsfig{figure=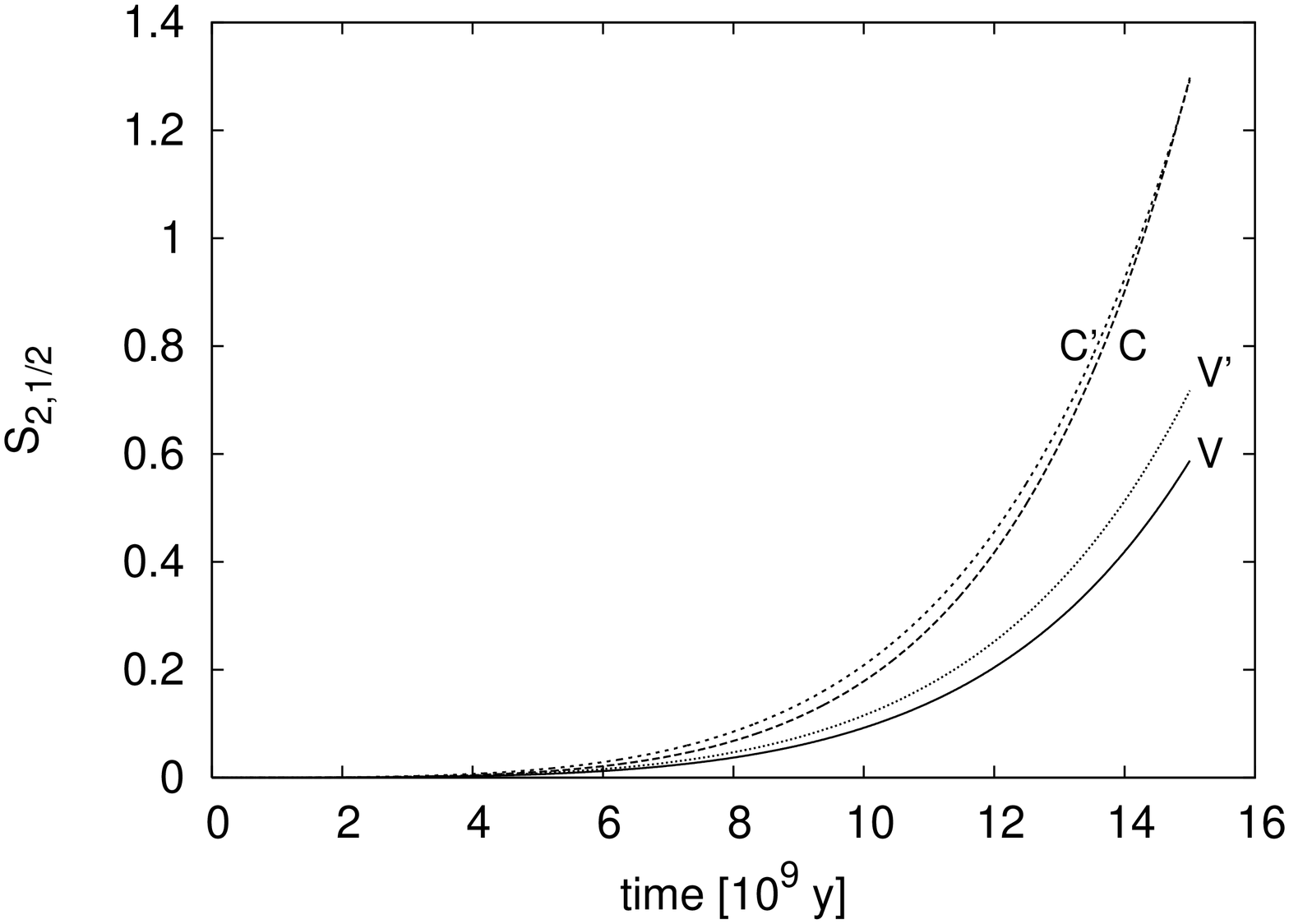, scale=0.24, angle=270}\figsubcap{a}}
  \hspace*{4pt}
  \parbox{2.1in}{\epsfig{file=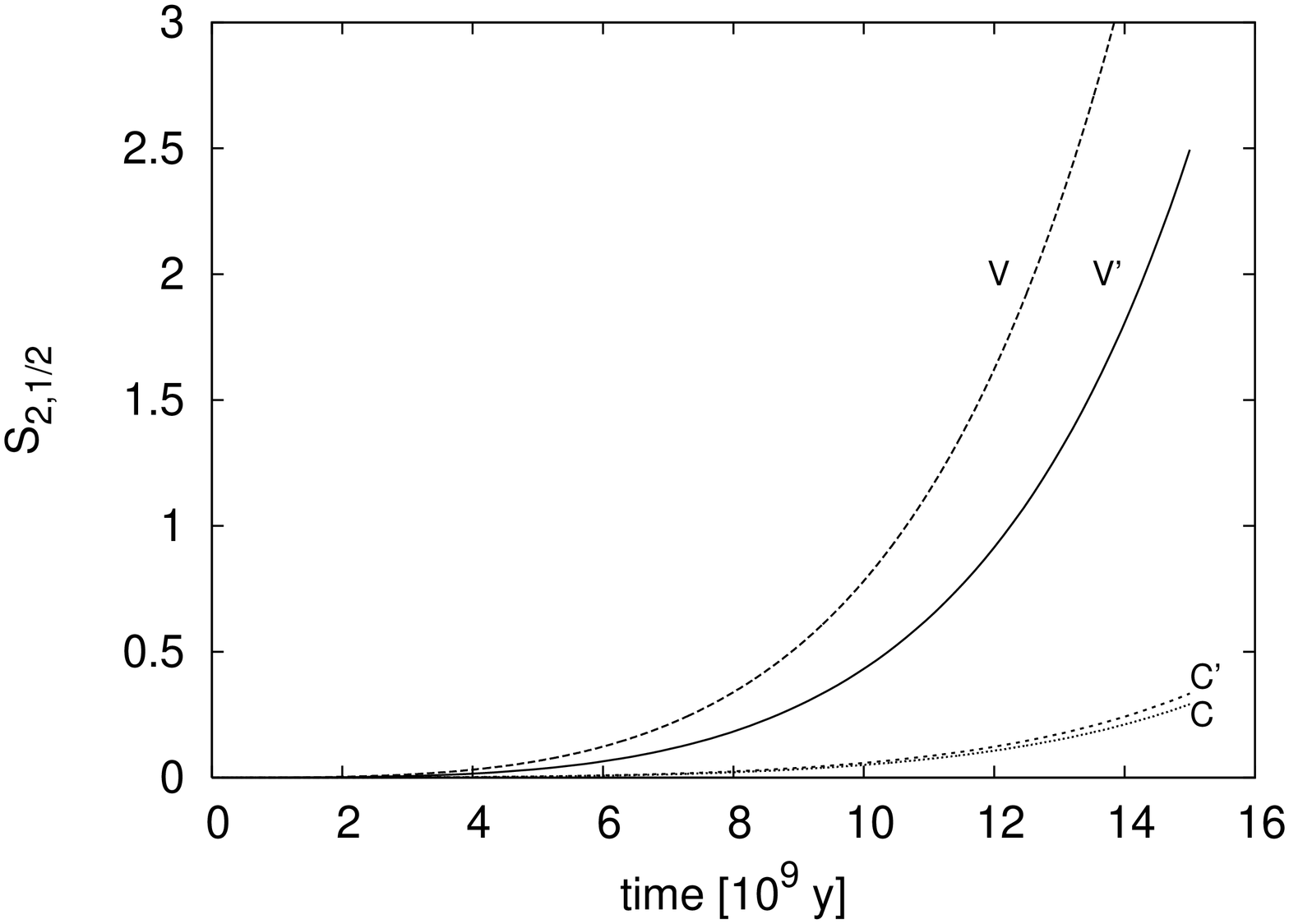, scale=0.24, angle=270}\figsubcap{b}}
  \caption{Comparison of $S_{2,1/2}$ for models with $\delta M<0$ (a) and
with $\delta M<0$ (b). C corresponds to a "supercluster" --- an overdense region, and V corresponds to a "void" --- underdense 
region. Primes denote models with $P' \ne 0 \ne S'$.
Since the value of $S_{\mathrm{IK}}$ depends on units,
the results presented here were normalised so they are now of order of unity.}%
  \label{fig7}
\end{center}
\end{figure}

Fig. \ref{fig7} presents the comparison of evolution of  $S_{2,1/2}$ for 
models 1--4.
The primed letters denote models of $S' \ne 0 \ne P', Q' \ne 0$.
As can be seen the evolution of $S_{2,1/2}$ for all these models is also comparable.
These results imply that the evolution in the quasispherical Szekeres model does not depend on the position of the dipole component. As long as the shape
 and density contrast  of the analysed models are similar,
such models evolve in a very similar way.

\section{The role of expansion}\label{expan}

The faster or slower evolution rate of the previously presented models
is reflected by their current expansion rate. As has been shown above, 
the evolution does not depend on a relative position of the dipole
component (evolution of models 1 and 3 is similar).
 Thus, let us focus on model 1 and model 2 only.

\begin{figure}%
\begin{center}
  \parbox{2.1in}{\epsfig{figure=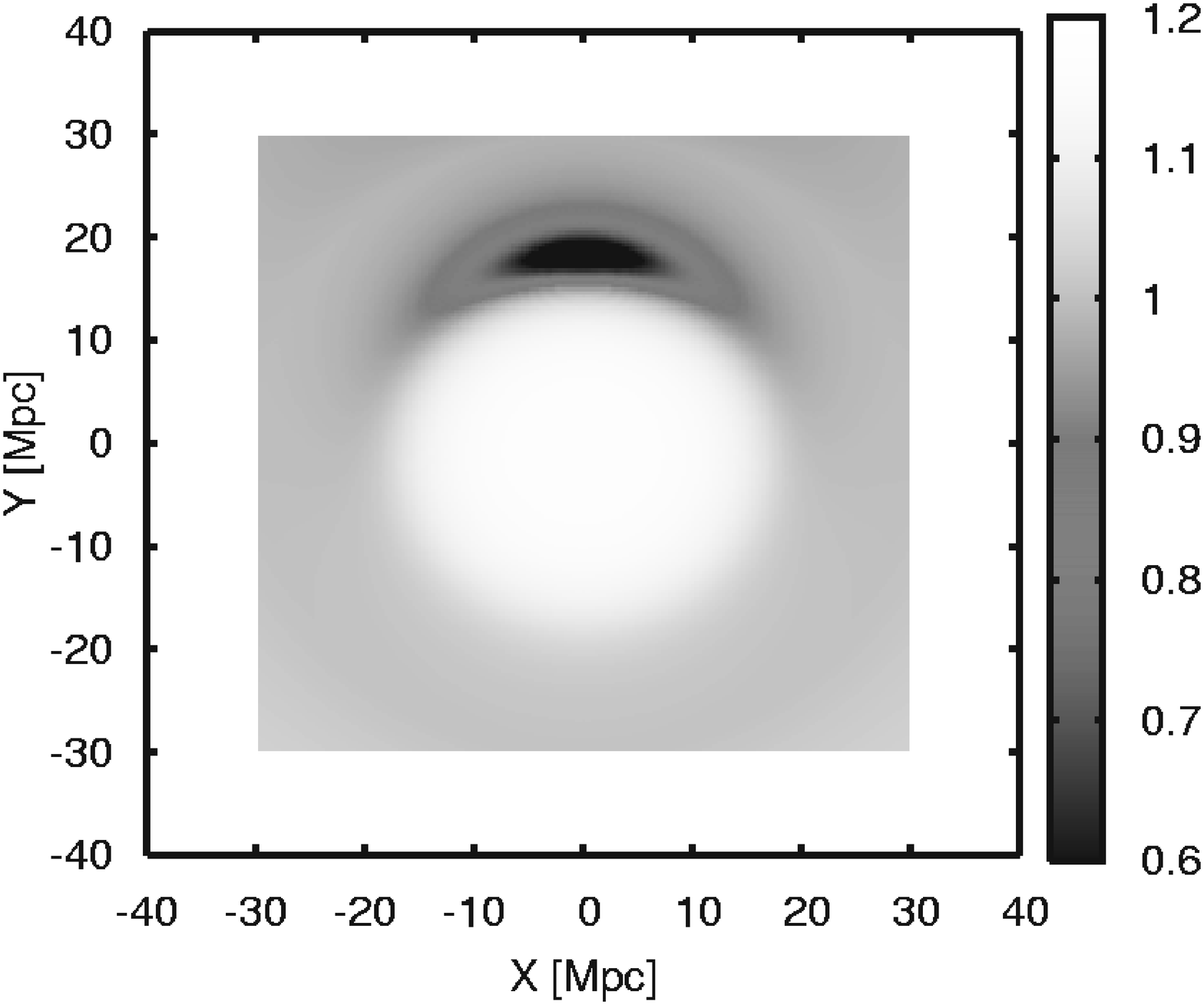, scale=0.2}\figsubcap{a}}
  \hspace*{4pt}
  \parbox{2.1in}{\epsfig{figure=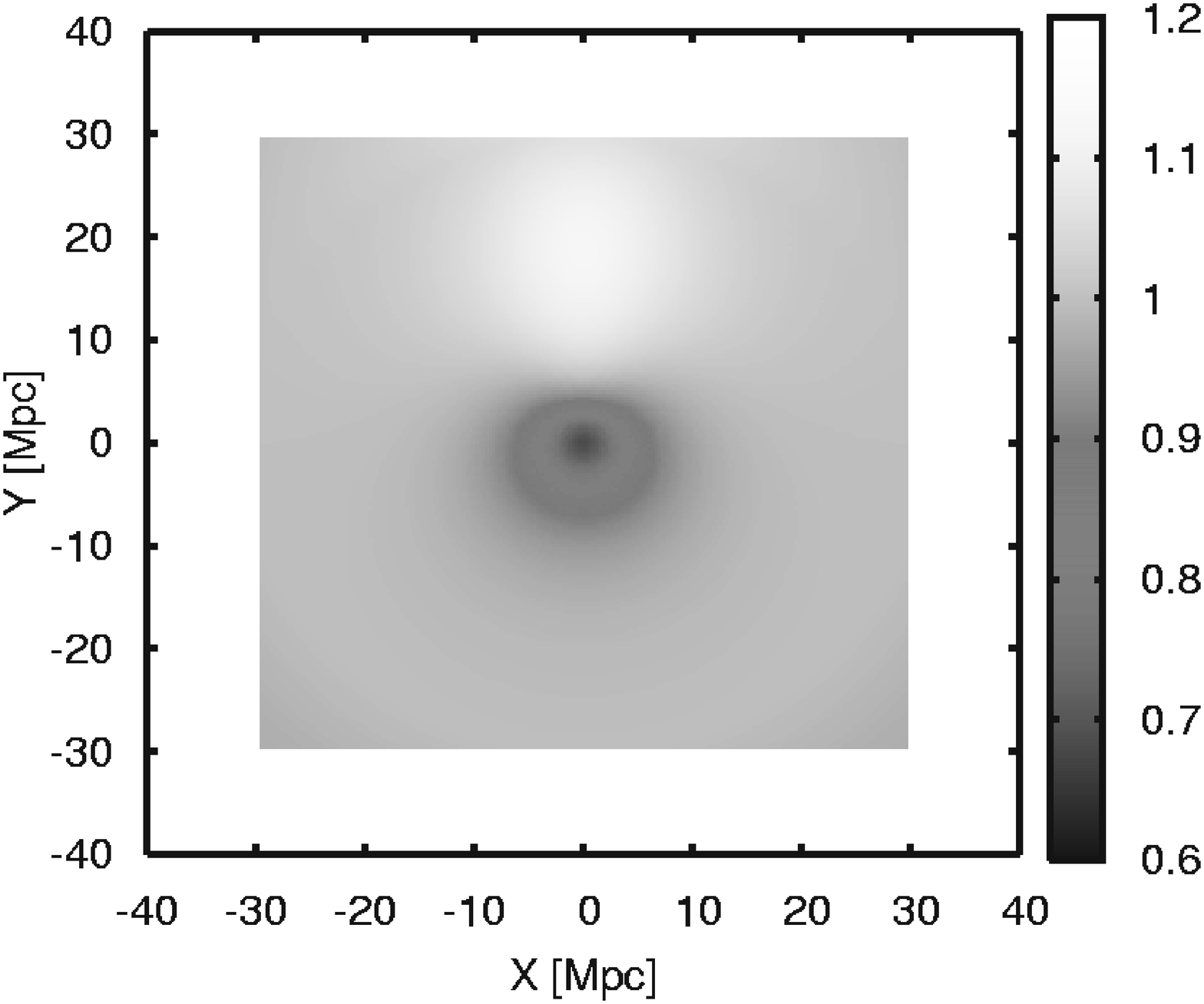, scale=0.2}\figsubcap{b}}
  \caption{The $\Theta_{\mathrm{SZ}} / \Theta_0$ ratio. (a) presents the
 ratio of model 1. (b) presents the ratio of model 2.}%
  \label{fig8}
\end{center}
\end{figure}

Fig. \ref{fig8} presents the ratio, $\Theta_{\mathrm{SZ}} / \Theta_0$, of the expansion parameter 
in the considered Szekeres models to the expansion parameter in 
the homogeneous background.
As can be seen, model 1, with $\delta M < 0$, has a larger amplitude of this ratio,
and the evolution of a supercluster in this model is much faster 
than in the corresponding Lema\^itre--Tolman model.
On the other hand, model 2 ($\delta M>0$) has smaller amplitude of the 
$\Theta_{\mathrm{SZ}} / \Theta_0 $ ratio and within model 2 the evolution of the
 density contrast inside the  void was much slower
than in the Lema\^itre--Tolman model.
So clearly the rate of the evolution is connected with the rate of the expansion. This conclusion is also supported by the continuity equation [Eq. (\ref{coneq})].
Higher mass in the perturbed region 
slows down the expansion rate --- 
this is a condition hindering the evolution of cosmic voids.
On the other hand, if the mass of perturbed
region is below the background mass, such region expands
much faster than the background, leading 
to the formation of the large underdense regions.
Such large voids enhance the formation
of large elongated overdensities (walls)
formed at the edges of voids.

\section{Conclusions}\label{concl}

The   galaxy redshift surveys show that  the Universe is patchy with various structures.
These structures include small voids among compact superclusters and large voids surrounded by large walls or long filaments.

The evolution of these cosmic structures 
in different environments in the quasispherical Szekeres model was investigated.
The Szekeres model is the most complex 
spatially inhomogeneous exact solutions 
of the Einstein field equations, and it is 
of great use in cosmology.
Since it is an exact
solution of Einstein's equations, it 
enables us to investigate the evolution of cosmic 
structures without such 
 approximations as linearity and small 
amplitude of density contrast.
Moreover, the Szekeres model is flexible enough to 
describe more than one structure.

Having investigated various models with two 
or three structures within one frame 
it may be concluded that 
the evolution of the cosmic structures 
depends on the environment.
In perturbed a region whose mass is below the
background mass the amplitude of the
expansion's fluctuations is large
and as can be seen from 
the continuity equation [Eq. (\ref{coneq})],
such conditions enhance the evolution of cosmic structures.

The analyses presented in this paper indicate that small voids 
among large overdense regions
do not evolve as fast as the large voids do.
This is because the expansion of the space is faster inside large voids than inside smaller voids.
Moreover, this higher expansion rate 
inside the large voids leads to the formation
of large and elongated structures such as walls and filaments
which emerge at the edges of these large voids.

\section*{Acknowledgments}
I would like to thank Andrzej Krasi\'nski and Charles Hellaby for their valuable comments and discusions concerning the Szekeres model.

\end{document}